# Discrete and parallel frequency-bin entanglement generation from quantum frequency comb

*Chi Lu[1], Xiaoyu Wu[1], Wenjun Wen[1] and Xiao-song Ma[1,2, ∗]*

[1]*National Laboratory of Solid-state Microstructures, School of Physics, College of Engineering and Applied Sciences, Collaborative Innovation Center of Advanced Microstructures, Jiangsu Physical Science Research Center, Nanjing University, Nanjing 210093, China*

[2]*Synergetic Innovation Center of Quantum Information and Quantum Physics, University of Science and Technology of China, Hefei, Anhui 230026, China*

∗*e-mails: Xiaosong.Ma@nju.edu.cn*

**Keywords**: quantum frequency combs, Hong-Ou-Mandel interference, quantum entanglement, frequency-bin entangled states

**Abstract:** Photons' frequency degree of freedom is promising to realize large-scale quantum information processing. Quantum frequency combs (QFCs) generated in integrated nonlinear microresonators can produce multiple frequency modes with narrow linewidth. Here, we utilize polarization-entangled QFCs to generate discrete frequency-bin entangled states. Fourteen pairs of polarization-entangled photons with different frequencies are simultaneously transformed into frequency-bin entangled states. The characteristic of frequency-bin entanglement is demonstrated by Hong-Ou-Mandel interference, which can be performed with single or multiple frequency pairs in parallel. Our work paves the way for harnessing large-scale frequency-bin entanglement and converting between different degrees of freedom in quantum information processing.

## 1. Introduction

Frequency-bin entanglement is crucial for quantum information processing [1,2]. Firstly, frequency-bin encoding is well-suited for single-mode optical fiber transmission, making it a promising candidate for realizing entanglement distribution in quantum networks using existing fiber infrastructure. Secondly, with the advancement of integrated photonics, it is now possible to generate a large number of correlated discrete frequency bins in parallel [3]. Thirdly, by using frequency-resolved techniques with mature dense wavelength-division multiplexing (DWDM) technology, one can realize straightforward filtering and control of quantum states encoded in frequency. Last but not least, recent achievements in high-speed electro-optical modulation with thin-film lithium niobate may help address previous challenges in manipulating frequency-bin quantum states [4].

With the development of integrated optics, 'quantum frequency comb' (QFC) can be produced with high brightness in a $\chi^{(3)}$ nonlinear optical micro-resonator through spontaneous four-wave mixing (SFWM) [3,5-15]. Rather than filtering after generation [16-18], photon pairs emitted from the microresonator automatically cover multiple discrete frequency modes, which avoids extra filtering loss. High quality (Q) factor ensures distinct separation between adjacent frequency modes and at the same time achieves higher SFWM efficiency due to cavity-enhanced effect. Such advantages make it a natural choice to generate multiple frequency-bin entangled states using QFC.

In this work, based on broadband polarization-entangled QFC in a Sagnac configuration [19], we manipulate more than 14 frequency pairs in parallel, deterministically transforming them into discrete frequency-bin entangled state through a hybrid quantum gate using free-space optical components [1]. We generate entangled photons with an average linewidth of about 190 MHz much smaller than the free spectral range (FSR) of about 99 GHz and the large frequency separations, ranging from nearly 200 GHz to more than 3 THz. By performing Hong-Ou-Mandel (HOM) interference [20,21] to characterize frequency-bin entangled states [1,22-24], the interference fringes exhibit biphoton coherence and have an oscillation, which can be harnessed in high temporal resolution in quantum metrology and sensing [25]. Moreover, by simultaneously measuring the interference fringe of multiple frequency pairs, the relative phase of QFC can be obtained. By adjusting the phase of the entangled state, we can arbitrarily alter the entanglement symmetry of the two-photon state. Finally, combined with the coincidence counts under computational basis, the restricted density matrix of the biphoton frequency-bin entangled state is reconstructed and thus verified.

Our work shows that, by employing the optical packaged silicon nitride chip within the Sagnac interferometer and a hybrid quantum gate, we generate the parallel two-photon frequency-bin entangled states formed by QFC and achieve their symmetry control. Based on the high-dimensional characteristic of QFC, our experiments further realize the generation and conversion between polarization entangled and frequency-bin entangled states with multiple frequency modes. We demonstrate a QFC source with high flexibility and versatility in application perspectives, which paves the way for harnessing large-scale entanglement of different degrees of freedom (DoFs) in quantum information processing, such as entanglement distribution in a quantum network.

## 2. Experimental Setup

First, as shown in **Figure 1(a)**, a chip-integrated silicon nitride microring resonator (MRR) is placed in a Sagnac interferometer to generate a broadband polarization entanglement quantum frequency comb (QFC) through the spontaneous four-wave mixing process [19]. The radius of our MRR is about $230\mu m$, corresponding to the 99GHz FSR at telecommunication wavelength. Photons are extracted from the cavity by point coupling between the MRR and a single bus waveguide, both of which have a cross-section of $0.8\mu m$ high and $1.6\mu m$ wide. We pump the source with a 193.5 THz (1549.3150 nm) continuous-wave pump laser at a low power level about 2.7mW on chip, generating a bi-photon bandwidth covering the telecom C-band [19,26]. The pump laser is edge-coupled into the chip from an ultra-high numerical aperture (UHNA7) fiber array with an insertion loss of 6dB, including both coupling and propagation loss. The MRR we use has an average FWHM of about 190.41MHz, an average Q-factor of about $1.03 \times 10^6$ and an FSR of about 99.03GHz, all of which are derived from the transmission spectrum as partially shown in **Figure 1(d)** and **(e)**. Correspondingly, the measured single-photon spectrum exhibits multiple discrete frequency modes with a bandwidth of about 190 MHz and a spectral spacing of about 99GHz, as shown in **Figure 1(f)**. We employ a C-band wavelength selective switch (WSS) as a programmable band-pass filter and a superconducting nanowire single-photon detector (SNSPD) to record the on-resonance and off-resonance photon counts with a step of 33GHz and a bandwidth of 20GHz.

In the Sagnac interferometer, the diagonally polarized pump photons are split on the input polarization beam splitter (PBS), propagating clockwise and counterclockwise, respectively. The horizontal components of the pump photons are rotated to vertical orientation by waveplates, while the vertical components remain unchanged. The MRR is bidirectionally pumped with the vertically polarized pump photons, corresponding to the TE00 mode in the waveguide. TE mode results in a lower optical loss than TM mode, leading to an enhanced Q-factor. Based on the SFWM process, pump photons are converted into vertically polarized photon pairs, which also propagate clockwise and counterclockwise, and are combined by the PBS. Then, the vertically polarized QFC and the horizontally polarized QFC rotated by HWP overlap in spectral and spatial modes[27,28]. The generated frequency-correlated polarization-entangled photons are close to the ideal quantum state [29]:

$$|\psi\rangle = \frac{1}{\sqrt{2}}\left(|H\rangle_s|H\rangle_i + e^{i\theta}|V\rangle_s|V\rangle_i\right)\otimes\frac{1}{\sqrt{M}}\left(\sum_{m=1}^{M}|\omega\rangle_{s,m}|\omega\rangle_{i,m}\right), \quad (1)$$

where $\theta$ can be compensated such that the $H/V$ components are in-phase, and $|\omega\rangle_{s(i),m}$ represents the state of the signal (idler) photon at frequency pair index $m$ $(m = 1,2,\dots,M)$. In our experiment, we use up to 14 frequency pairs, with a minimum separation of 0.396 THz and a maximum separation of 2.97 THz for the center frequency of the signal and the idler modes.

The experimental setup for generating discrete frequency-bin entanglement is shown in **Figure 1(b)**. A C-band WSS is connected to the polarization-entangled QFC source via a single-mode fiber. It demultiplexes low-frequency modes (signal photons) and high-frequency modes (idler photons) into different spatial modes. The separated signal photon and idler photons are then sent to the two inputs (port-1 and port-2) of a PBS after passing through fiber polarization controllers (FPC) respectively. The FPC on each path is used to align the polarization axes of the polarization-entangled QFC source with those of the PBS. The horizontally polarized $|\omega\rangle_s$ and vertically polarized $|\omega\rangle_i$ are routed by PBS to output port-3, and the vertically polarized $|\omega\rangle_s$ and horizontally polarized $|\omega\rangle_i$ are routed to output port-4. This maps the existing polarization entanglement to the frequency-bin DoF. Polarizers at 45° behind the two outputs are used to erase polarization information and to project the photons onto diagonal polarization. A combination of a quarter-wave plate, a half-wave plate and a quarter-wave plate (QWP-HWP-QWP), with both QWPs oriented at 45°, are inserted to arbitrarily control the phase of each frequency pair. The generated bi-photon state, consisting of one pair of frequency bins, can then be expressed as [1]:

$$|\psi\rangle_m = \frac{1}{\sqrt{2M}}\left[(|\omega\rangle_{s,m}|\omega\rangle_{i,m} + e^{i\theta}|\omega\rangle_{i,m}|\omega\rangle_{s,m})\right]. \tag{2}$$

This state represents a discrete frequency-bin entangled state with multiple frequency bins in parallel. We utilize the WSS to implement the selection of frequency bins, either a single pair or multiple pairs.

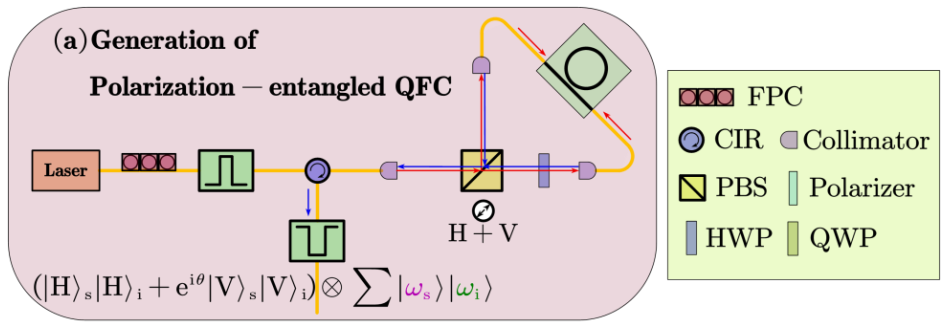

(a) Generation of Polarization−entangled QFC

$(|H\rangle_s|H\rangle_i + e^{i\theta}|V\rangle_s|V\rangle_i) \otimes \sum |\omega_s\rangle|\omega_i\rangle$

FPC, CIR, Collimator, PBS, Polarizer, HWP, QWP

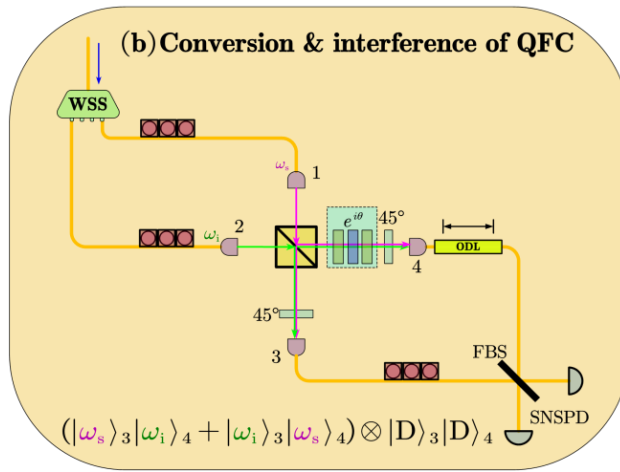

(b) Conversion & interference of QFC

$(|\omega_s\rangle_3|\omega_i\rangle_4 + |\omega_i\rangle_3|\omega_s\rangle_4) \otimes |D\rangle_3|D\rangle_4$

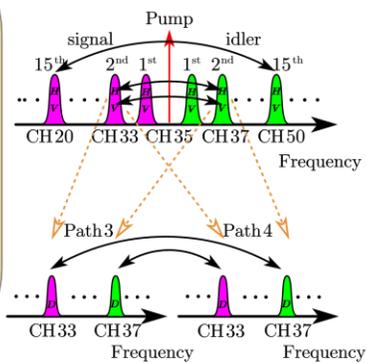

(c) Spectrum of QFC

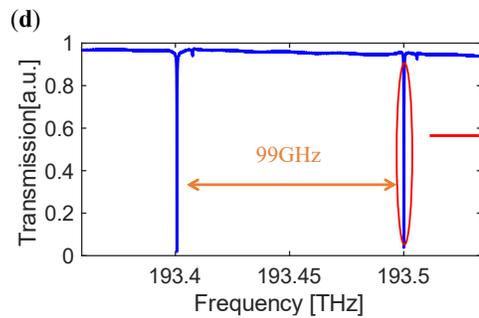

(d)

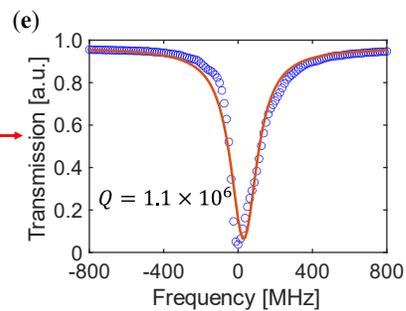

(e) $Q = 1.1 \times 10^6$

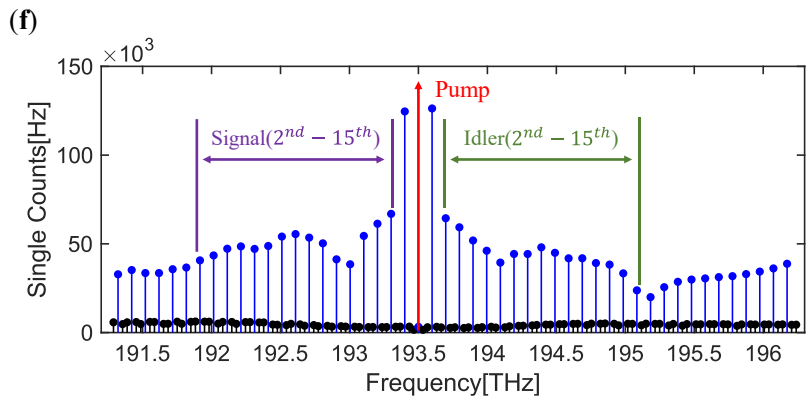

(f) Signal($2^{nd} - 15^{th}$), Pump, Idler($2^{nd} - 15^{th}$)

**Figure 1.** Experimental setup for the generation of discrete and parallel frequency-bin entangled states. (a) Polarization-entangled quantum frequency comb (QFC) is implemented with a Sagnac configuration. QFC is generated in the silicon nitride micro-ring resonator. (b) Signal (at low frequency) and idler (at high frequency) photons are divided into two paths and converted into frequency-bin DoF entanglement by a PBS and a 45° polarizer. Two-photon interference at FBS is detected by SNSPD. (c) Schematics of QFC's spectrum. More than 14 frequency pairs are symmetrically distributed on both sides of the pump frequency(193.5THz) with about 99GHz FSR. FPC, fiber polarization controller; CIR, circulator; PBS, polarization beam splitter; HWP, half-wave plate; QWP, quarter-wave plate; WSS, wavelength selective switch; ODL, optical delay line; FBS, fiber beam splitter; SNSPD, superconducting nanowire single-photon detector. (d)Transmission spectrum near the pump frequency of 193.5THz with an FSR of 99GHz, which agrees with the designed 100GHz. (e)Resonance dip at the pump frequency with a Q-factor of $1.1 \times 10^6$. (f)Single-photon spectrum covering the C-band. On-resonance lines are blue and off-resonance lines are black.

## 3. Experimental Results

To characterize frequency-bin entanglement, we perform nonclassical Hong-Ou-Mandel (HOM) interference measurements on non-degenerate signal-idler photon pairs with different frequency detunings.

A photon emitted from the PBS passes through a tunable optical delay line (ODL) and coincides with another photon passing through an FPC on a 50:50 fiber beam splitter (FBS). The FPC is used to match the polarization of two photons. The relative arrival times of two photons are adjusted using the fiber ODL to manipulate the temporal degrees of freedom, thereby introducing distinguishability. The signal and idler photons interfere at the FBS, and the coincidence counts exbibit a sinusoidal oscillation, expressed as [17,30,31]:

$$p_c(\tau) = \frac{1}{2}\{1 - V \cos[(\omega_i - \omega_s)\tau + \phi]\} \cdot \text{Re} \int d\Omega |f(\Omega)|^2 e^{2i\Omega t}, \qquad (3)$$

where $\tau$ is the relative delay between the signal photon and the idler photon, $V$ is the visibility of the interference fringe, and $\phi$ is the phase of the quantum beat frequency. The beat pattern exhibits an envelope that is determined by the transmission lineshape of the MRR. Here, $f(\Omega) = \frac{A}{\Omega^2 + (\sigma/2)^2}$ is the Lorentz lineshape, where $A$ is the normalization coefficient and $\sigma$ is the linewidth. By fitting the measured interference fringe, we obtain the visibility $V$ and phase $\phi$, respectively.

We program the WSS as a two-channel bandpass filter to separate signal photons from idler photons. **Figure 2(a)** shows the normalized coincidence count rate as a function of the relative delay time for the 2$^{nd}$ frequency pair (ITU-CH33 and CH37, with a central wavelength of 193.3 THz and 193.7 THz, respectively). The scanning range of ODL is from 0 to 2.4 ns with a resolution of 2 ps. The frequency detuning between the signal photon and idler photon is 4 times the FSR, that is, 396 GHz. The corresponding center frequencies of the two WSS channels are set to 193.302 THz and 193.698 THz, respectively, with a channel bandwidth of 20 GHz. Due to limitations in ODL performance (scan range up to 2.5 ns) and large coherence length of the photon pairs, we obtain a partial HOM interference curve, whose envelope agrees well with the theoretical envelope determined by the Lorentz line shape of the cavity resonance. The width of the HOM envelope is approximately 8 ns, determined by the single-photon bandwidth of about 190MHz, which is much larger than the oscillation period determined by frequency detuning. We further use ODL to scan with a higher precision of 0.1 ps to obtain finer interference curves. The results are presented in **Figure 2(b)** and **2(c)**, respectively. The visibility of the interference fringe decreases from 78.62 $\pm$ 2.63% in **Figure 2(b)** to 33.45 $\pm$ 4.01% in **Figure 2(c)**, as the temporal distinguishability of two photons increases.

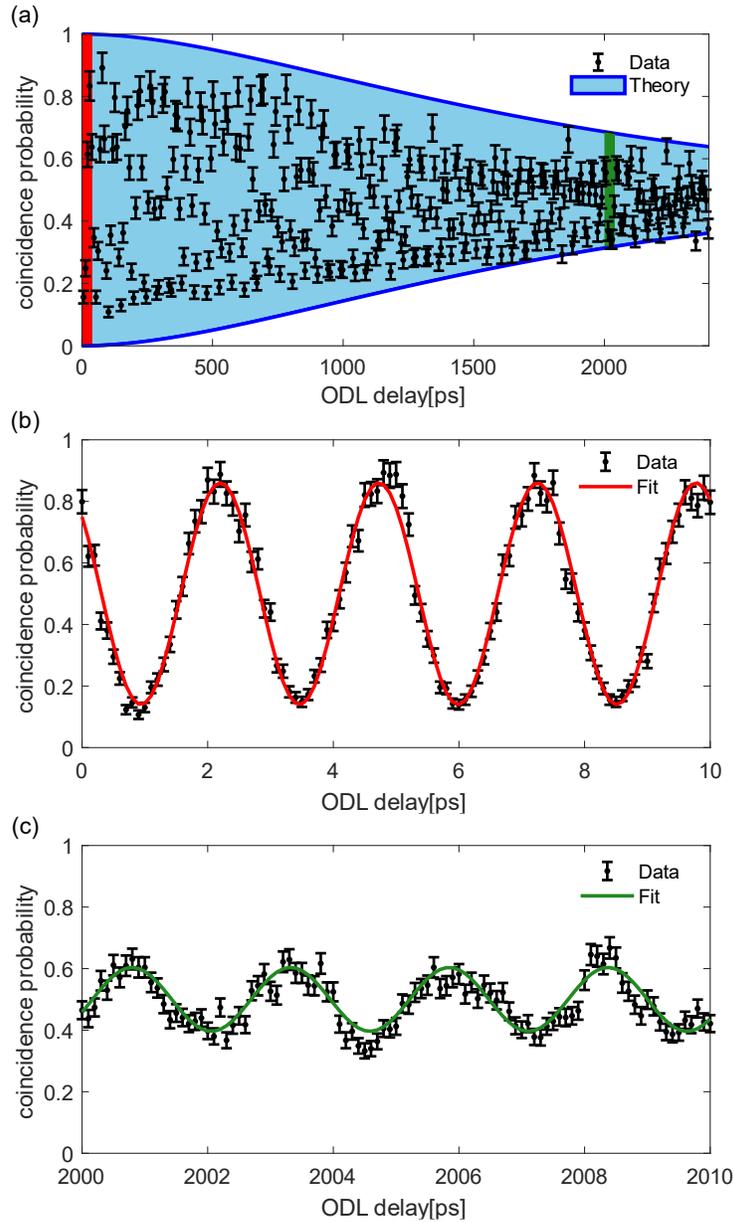

**Figure 2.** Two-photon interference of $2^{nd}$ frequency pair. The delay set by ODL is not equivalent to the two-photon relative path delay. The partial envelope curve is shown in (a) with experimental results and theoretical predictions based on cavity resonance measurements. Finer interference curves are plotted in (b) and (c) near 0ns and 2ns respectively with experimental and fitting results.

We further measure the HOM interference curve of the $5^{th}$, $10^{th}$, and $15^{th}$ pairs of frequency channels by changing the center frequencies of the WSS channels. The corresponding frequency detunings are about 0.99 THz, 1.98 THz and 2.97 THz, respectively. The oscillation periods of the interference fringes shown in **Figure 3(a-c)** are 1.01ps, 0.51ps, and 0.34ps respectively, consistent with the detuning of the frequency pair we use. The three interference fringes exhibit visibilities of $78.43 \pm 2.37\%, 86.89 \pm 1.55\%$, and $86.83 \pm 2.22\%$, respectively, indicating the frequency-bin entanglement for the respective pair of the conjugate frequency channels.

Furthermore, we program the WSS by increasing its passband width, as a multi-channel bandpass filter. This allows multiple frequency channel pairs to be multiplexed simultaneously. Theoretically, the result of HOM interference when multiplexing multiple frequency channel pairs is the sum of the individual interference curves [30-32], expressed as:

$$p_c(\tau) = \sum_{m=1}^{M} \frac{1}{2} \{1 - V_m \cos[(\omega_{i,m} - \omega_{s,m})\tau + \phi_m]\} \cdot \text{Re} \int d\Omega |f(\Omega)|^2 e^{2i\Omega t}. \quad (4)$$

**Figure 3(d-f)** show the interference results when 4 pairs ($2^{nd}$ to $5^{th}$ pair), 9 pairs ($2^{nd}$ to $10^{th}$ pair) and 14 pairs ($2^{nd}$ to $15^{th}$ pair) of frequency channels are multiplexed respectively. The frequency separation of the $n^{th}$ photon pair is 2n times the FSR and phases of individual interference curves are consistent, which leads to the revival of coincidence dips. The period of recurrence is about 5ps, corresponding to the detuning of the $1^{st}$ frequency pair, which is the twice of FSR. The dip becomes sharper as more frequency pairs are multiplexed, which indicates the increasing maximum frequency separation of the frequency-bin entangled states. However, as more frequency pairs are multiplexed, channels belonging to different frequency pairs will contribute to accidental coincidences, reducing visibility. These results cannot reflect the relative phase between different frequency pairs [31], and the full characterization of high-dimensional frequency-bin entanglement requires the necessary projection measurement at the frequency DoF [33-37].

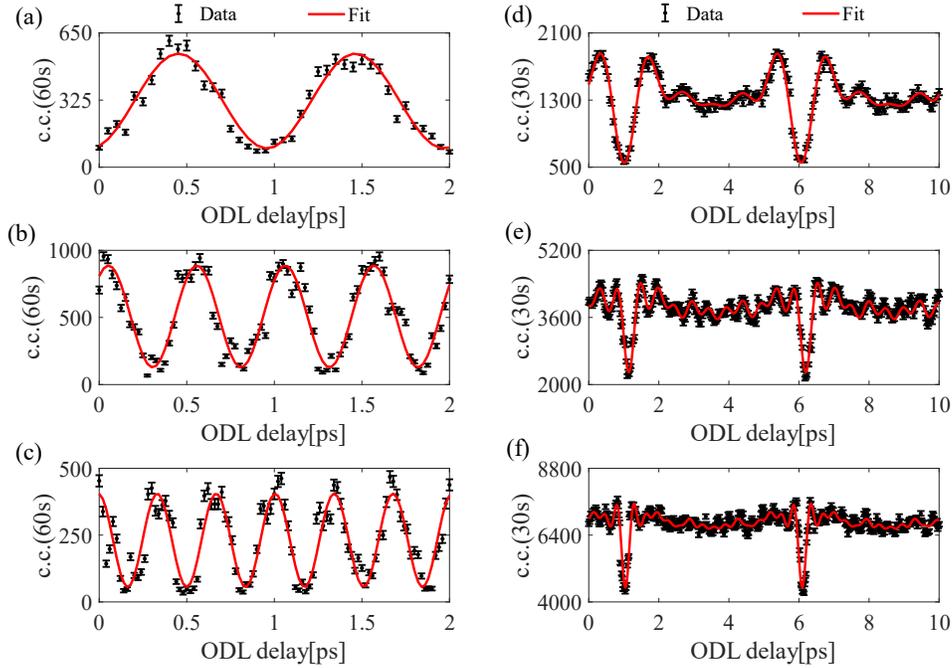

**Figure 3.** Two-photon interference of different frequency pairs. (a-c) Coincidence counts in 60s of the 5[th], 10[th] and 15[th] frequency pair (193.0THz-194.0THz, 192.5THz-194.5THz and 192THz-195THz). (d-f) Coincidence counts in 30s of the 2-5[th], 2-10[th] and 2-15[th] frequency pairs. Red lines are fitting results considering the accidental coincidence counts.

The phase $\theta$ between $|\omega\rangle_{s,m}|\omega\rangle_{i,m}$ and $|\omega\rangle_{i,m}|\omega\rangle_{s,m}$ in the frequency-bin entangled state can be adjusted using the "QWP-HWP-QWP" setup. The range of $\theta$ is from 0 to $2\pi$, which changes the exchange symmetry of the entangled state. Correspondingly, the peaks and valleys of the HOM fringes will alternate with the phase change. **Figure 4(a)** shows the interference fringes of the 2[nd] pair of frequency channels when the relative phases are 0, $\pi/2$, $\pi$, and $3\pi/2$, respectively, representing four discrete frequency-bin entanglement states close to the maximum entanglement states, each of which exhibits an interference visibility over 75%. We also measured the interference fringes under different relative phases by multiplexing multiple pairs of frequency channels. **Figure 4(b)** shows the result when 4 pairs (2[nd] to 5[th] pair) are multiplexed. This result clearly shows the photons' spatial bunching and anti-bunching with different relative phases, revealing the spectral symmetry and anti-symmetry of the entangled photon pairs [17,38]. It can also be seen that the relative phase does not change for different frequency pairs, which indicates the ability of parallel control over high-dimensional frequency-bin entangled states formed by QFC.

Written in the computational basis, the density matrix of the biphoton frequency-bin entangled state is [1]:

$$\rho = \begin{pmatrix} 0 & 0 & 0 & 0 \\ 0 & p & \frac{V}{2}e^{-i\phi} & 0 \\ 0 & \frac{V}{2}e^{i\phi} & 1-p & 0 \\ 0 & 0 & 0 & 0 \end{pmatrix}$$

where $p$ is the balance parameter, $V$ is the visibility and $\phi$ is the phase of the quantum beat. The parameter $p$ can be obtained from the coincidence counts in the computational basis ($5914 \pm 77$ and $2527 \pm 50$ for $|\omega\rangle_{s,2}|\omega\rangle_{i,2}$ and $|\omega\rangle_{i,2}|\omega\rangle_{s,2}$), and its value is $0.701 \pm 0.005$. It is not equal to 0.5 due to imbalances in fiber losses and coupling efficiency for the different frequency bins. Combined with the visibility $V$ ($V = 77.13 \pm 1.93\%$, obtained from the quantum beat of the 2nd pair of frequency channels) and the phase $\phi$ ($\phi = -0.1168 \pm 0.1094$ rad, fitted from that of multiple frequency pairs), we can reconstruct the reduced density matrix [1], with its real and imaginary parts shown in **Figure 5(a)** and **5(b)** respectively. Its fidelity relative to the target maximum entangled state is $88.30 \pm 1.15\%$.

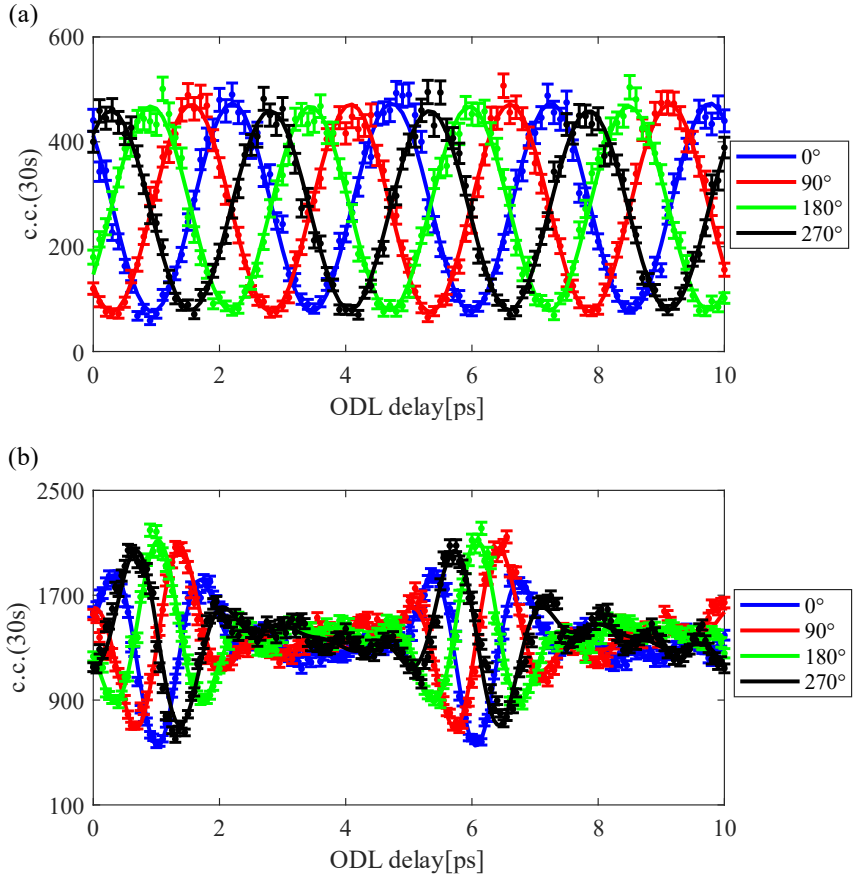

**Figure 4.** Hong-Ou-Mandel interference of frequency-bin entangled photons with the relative phase $\phi = 0°, 90°, 180°\ and\ 270°$. (a) Coincidence counts in 30s of the 2$^{nd}$ frequency channel. (b) Coincidence counts in 30s of the 2-5$^{th}$ frequency channels.

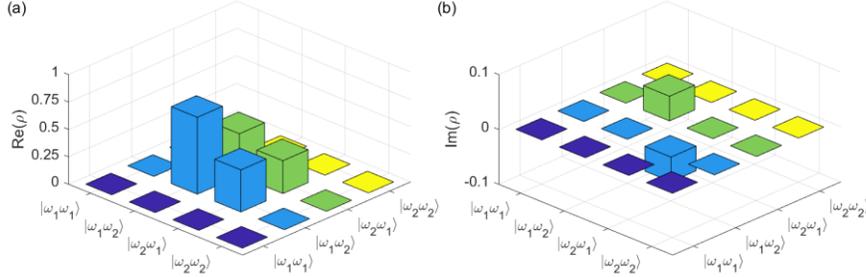

**Figure 5.** The density matrix of two-photon frequency-bin entanglement state for the 2$^{nd}$ frequency channel (193.3THz-193.7THz). (a) The real part of the density matrix. (b) The imaginary part of the density matrix.

## 4. Conclusion

In conclusion, we obtain a multi-mode discrete frequency-bin entangled state converted from a polarization-entangled biphoton QFC. We show that our integrated silicon nitride MRR, which provides a large ratio of FSR to single-photon bandwidth, guarantees the well-separation between different frequency modes. Frequency DoF can transmit information over long distances with its immunity to birefringence and can be scaled to higher dimensions easily due to the inherent properties of the quantum frequency comb. We employ HOM interference to demonstrate every two-dimensional frequency-bin entangle state constituted by a pair of correlated frequency channels. Our results indicate the excellent properties of the QFC source based on silicon nitride integrated photonics platforms in generating entanglement with different DoFs over multiple frequency modes, which offers more versatility for entanglement distribution in a fully connected multi-user quantum network [26,39-41].

**Acknowledgments**

This research was supported by the National Key Research and Development Program of China (Grants Nos. 2022YFE0137000, 2019YFA0308704), the Natural Science Foundation of Jiangsu Province (Grants Nos. BK20240006, BK20233001), the Leading-Edge Technology Program of Jiangsu Natural Science Foundation (Grant No. BK20192001), the Innovation Program for Quantum Science and Technology (Grants Nos. 2021ZD0300700 and

2021ZD0301500), and the Fundamental Research Funds for the Central Universities (Grants Nos. 2024300324).